# Bipolar nanochannels: The effects of an electro-osmotic instability. Part I: Steady-state response

Ramadan Abu-Rjal and Yoav Green*
Department of Mechanical Engineering, Ben Gurion University of the Negev, Beer-Sheva 8410501, Israel
* Corresponding author. E-mail: yoavgreen@bgu.ac.il

Abstract
The steady-state current-voltage response of ion-selective systems varies as the number of ion-selective components is varied. For the highly investigated unipolar system, including only one ion-selective component, it has been shown that above a supercritical voltage, an electroosmotic instability is triggered, leading to over-limiting currents. In contrast, the effects of this instability on the current-voltage response of the second most common system of a bipolar system, including two oppositely charged permselective regions, have yet to be reported. Using simulations, we investigate the steady-state electrical response of bipolar systems as we vary the ratio of the charge within the two oppositely charged regions. The responses are divided into those with an internal symmetry related to the surface charge and those without. In contrast to the unipolar systems, bipolar systems with the internal symmetry do not exhibit overlimiting currents, and their steady-state response is identical to the convectionless steady-state response. In contrast, the systems without the internal symmetry exhibit much more complicated behavior. For positive voltages, they have overlimiting currents, while for negative voltages, they do not have over-limiting currents. Our findings contribute to a more profound understanding of the behavior of the current-voltage response in bipolar systems.

> **Impact Statement.** Permselective nanoporous materials are ubiquitous in desalination, energy harvesting, and bio-sensing systems. Of particular importance are bipolar membranes and nanochannels that are comprised of two oppositely charged permselective regions. While a plethora of experimental works have characterized the electrical response of these systems, a fundamental understanding of the underlying physics determining the response is still missing. To address this knowledge gap, we have systematically simulated different bipolar nanofluidics systems subject to varying potential drops and characterized their electrical response to reveal signatures that are unique to every system. Our findings contribute to a more profound understanding of the various control parameters and mechanisms that determine the time transient dynamics and the steady-state current-voltage response in bipolar systems and provide a valuable tool for interpreting experimental and numerical data of such systems. The insights from this work can be used to improve the design of fabricated bipolar devices.

## 1. Introduction

Ion-selective mediums (e.g., ion-exchange membranes, nanochannel, and/or nanopores) have been the focus of intensive research owing to their significant roles in numerous applications, such as desalination (Nikonenko et al., 2014; Tunuguntla et al., 2017; Marbach & Bocquet, 2019), energy harvesting (Siria et al., 2013; Bocquet, 2020; Kavokine, Netz, & Bocquet, 2021; Wang et al., 2023), biomolecule sensing (Slouka, Senapati, & Chang, 2014; Vlassiouk, Kozel, & Siwy, 2009), and fluidic based electrical circuits (Vlassiouk, Smirnov, & Siwy, 2008; Lucas & Siwy, 2020; Noy, Li, & Darling, 2023). An ion-selective medium's ability to undertake such a wide range of tasks stems from its surface charge, whereby counterions (ions with opposite charge to that of the surface charge) have preferential transport over the coions (ions with a similar charge to that of the surface charge).

**Figure 1** shows a prototypical setup used for the aforementioned applications. The system is comprised of either three or four layers – the exact number of layers and their roles will be discussed shortly. At the two ends of the system are two electrodes under an applied voltage drop of $V$, which measure a current, $I$ (or vice versa, an applied current and measured voltage). This current-voltage response, $I-V$, which is the primary characterizer of ion transport across ion-selective materials, depends on the layout of the system (i.e., three or four layers) and thus will be discussed shortly.

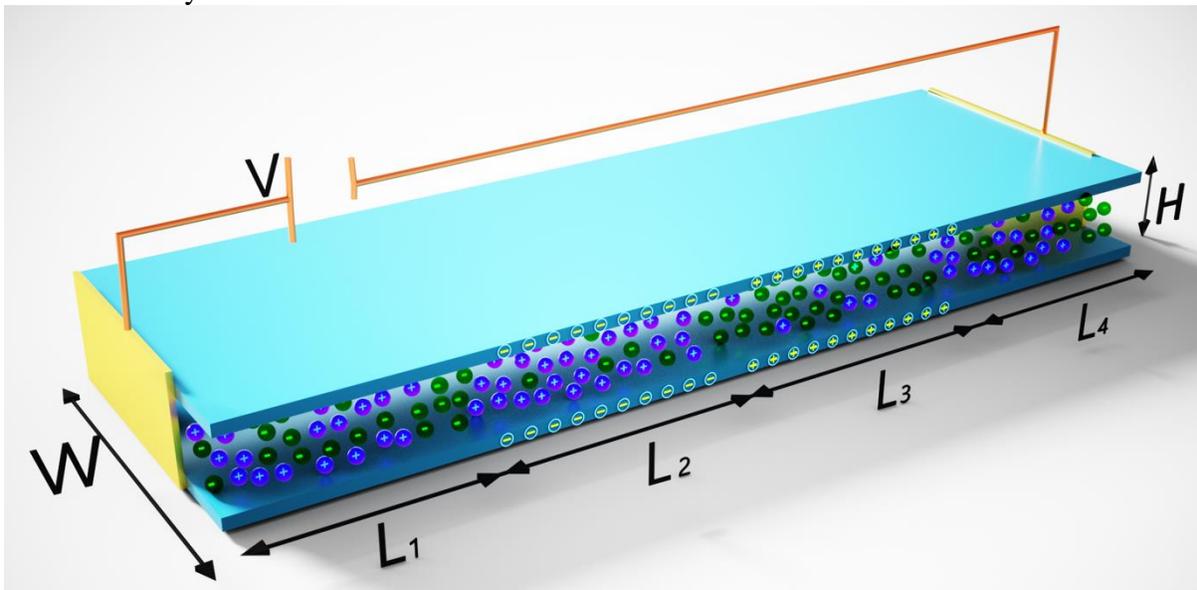

*Figure 1. Schematic of a three-dimensional four-layered system comprised of two diffusion layers connected by two permselective mediums under an applied voltage drop, $V$. The length of each of the four regions ($k=1,2,3,4$) is given by $L_k$, the height is $H$ and the width is $W$. The two outer regions are uncharged such that the concentrations of the positive ions (purple spheres) and the negative ions (green spheres) are the same. The two middle regions are charged with either a negative or positive surface charge density, leading to a surplus of counterions over coions. In the negatively charged region, the positive ions are the counterions, while in the positively charged regions, the negative ions are the counterions.*

In both the three-layer scenario and the four-layer scenario, adjacent to the electrodes, are two regions that can be considered to be independent of the surface charge densities. Consequently, as a result of negligible surface charge effects, the counterion and coion concentrations are virtually identical, such that these two regions are electroneutral. Depending on their geometry, these regions are often called reservoirs, microchannels, or simply the "diffusion-layer". We will use the latter term, which is the most general.

In between the diffusion layers is the all-important highly-charged ion-selective material, which can take different configurations. The two simplest configurations are what we term unipolar and bipolar nanochannels, corresponding to three-layers and four-layers, respectively. A unipolar nanochannel is a nanochannel that has only one charged region, which is either negatively charged or positively charged. In contrast, the bipolar nanochannel has two charged regions – one negative and one positive. This double-charged region introduces an additional "internal degree of symmetry" into the system. As a result, the responses of unipolar and bipolar systems behave drastically differently. More complicated configurations that include three or more charged regions are possible (Mádai, Matejczyk, Dallos, Valiskó, & Boda, 2018; Noy et al., 2023) but will not be addressed here.

In this two-part work, we will consider both the time-transient and steady-state responses of a bipolar nanofluidic system subject to a supercritical voltage drop that leads to an electro-osmotic instability (EOI) at the interface of the diffusion layer and the permselective material. To that end, we will utilize previous understandings of the time-transient and steady-state responses of unipolar systems, both without and with electro-osmotic flow (EOF), as well as the time-transient response of a bipolar system without EOF. When considering both time-transient and steady-state responses, it is natural to first consider the time-transient response that leads up to the steady-state response. However, we have found it best first to elaborate and discuss the steady-state response, which is characterized by the more common and intuitive current-voltage response. Thereafter, we will rationalize the results by considering the more involved time-transient response. Thus, we have divided this work into two parts: Part I – steady-state response, and Part II – time transient response. For the sake of brevity, we will refer to these works as Part I and Part II.

In the following two-part work, to reduce the computational costs, we will consider a two-dimensional (2D) system, which will be the top view of **Figure 1**; hence, we have illustrated all four regions to be of the same height and width. In general, this is not the case as realistic systems are three-dimensional (3D) and heterogeneous, whereby the heights and widths of the diffusion layers are substantially larger than those of the ion-selective regions (Sebastian & Green, 2023). Several works have shown that the transition from 3D to 2D does not change the robustness or generality of the results (Demekhin, Nikitin, & Shelistov, 2014; Druzgalski & Mani, 2016; Pham et al., 2016; Kang & Kwak, 2020), but it does make all numerical computations tractable. Also, in general, the ion-selective material in the center is a nanoporous membrane, which has a complicated geometry that is not as simple as the nanochannel portrayed in **Figure 1**. Since a single nanochannel and complicated membrane share the same ion-selective property, this reduction doesn't change the governing physics. Further, since we will consider a 2D system, the only important property that must be accounted for is the surplus of counterions due to the surface charge density. Thereafter, nanoporous materials and nanochannels behave precisely the same (Rubinstein et al., 2008; Yossifon & Chang, 2008; Sebastian & Green, 2023), thus, we can use either terminology (nanoporous materials and nanochannels) interchangeably.

The paper is structured as follows. Section 2 reviews the current-voltage response of unipolar and bipolar systems. Section 3 presents the 2D time-dependent model of the four-layered system shown in **Figure 1** that is solved numerically. Section 4 reports and discusses the steady-state response. Here and in Part II (Abu-Rjal & Green, 2024), when discussing bipolar systems, we make an artificial division into two scenarios: the "ideal" bipolar system and the "non-ideal" bipolar system. The definition of "ideal" vs. "non-ideal" will be discussed in Sec. 4. Importantly, while we will show that the "ideal" bipolar system has completely different characteristics compared to that of the unipolar system, the "non-ideal" bipolar system will share characteristics of both systems. Section 5 will include a discussion and summary of our results.

## 2. Steady-state current-voltage responses

In the following, we will describe the differences between the $I-V$ curves of unipolar and bipolar systems without and with the effects of an electro-osmotic flow (EOF). To remove the dependency of the current, $I$, on the width, $W$, and height, $H$, of the system, we will consider the current density, $i$. **Figure 2** presents three $i-V$ curves of the three known scenarios: unipolar without and with EOF, and bipolar without EOF. This work focuses on delineating the unknown curve of a bipolar system with EOF.

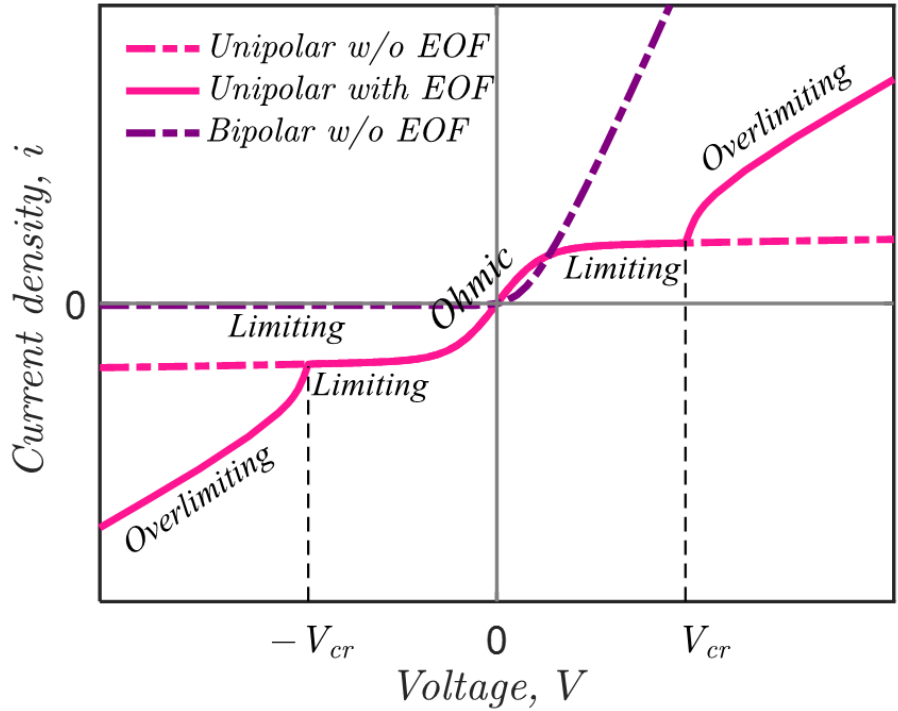

*Figure 2: Schematic of the current density−voltage ($i-V$) response curve for three different systems: unipolar systems, without and with EOF, and a bipolar system without EOF. The $i-V$ of the unipolar system is comprised of three distinct regions: the Ohmic (i.e., linear response regime), the limiting current density, and the overlimiting current density responses. The bipolar response has a limiting current for negative voltages and currents of unconstrained values for positive voltages.*

### 2.1 *Current-voltage response of unipolar systems without EOF*

The simplest and most investigated scenario is that of a unipolar system without EOF (dashed pink line in **Figure 2**). Here, it can be observed that the $i-V$ has two distinct regions. At low voltages and low currents, the response is linear and is characterized by the Ohmic response (Levich, 1962). At high voltages, when the concentration near the interface approaches zero, the current 'almost' saturates to a limiting value known as the limiting current, $i_{\text{lim}}$, which is determined by a diffusion process. A complete saturation supposedly occurs (i.e., a zero slope) if one assumes that the electric double layer (EDL) is substantially smaller than the diffusion layer and that the EDL has an equilibrium structure. However, Rubinstein and Shtilman (1979), and later Yariv (2009), showed that at these high voltages, the EDL loses its quasi-equilibrium-like structure, and a nonequilibrium extended space charge (ESC) layer is formed that increases the current slightly above the predicted limiting current. These currents were later termed by Yariv (2009) to be above limiting currents (which are independent of EOF) and are not to be confused with over-limiting currents (OLC), $i_{OLC}$, which will now be discussed and strongly dependent on EOF [see Green (2020) – for a discussion on the subtleties between the over-limiting and above-limiting terminology].

## 2.2 *Current-voltage response of unipolar systems with EOF*

While OLCs had been experimentally measured since the middle of the 20[th] century, they remained a mystery until the seminal works of Rubinstein and Zaltzman who showed that above a critical voltage, $V_{cr}$, an electro-osmotic instability (EOI) forms at the interface of the diffusion layer with the unipolar material (Rubinstein & Zaltzman, 2000; Zaltzman & Rubinstein, 2007) – this is the solid pink line in **Figure 2**). The appearance of EOI is due to the ESC losing stability due to transverse perturbations (i.e., EOI cannot appear in a 1D system or 2D system without EOF). Once the EOI has been initiated, the electrolyte, which until now has been dominated by a diffusion process, is now dominated by a more effective convective mixing process, which increases the current to over-limiting currents. The experimental verification of EOI as the source of OLCs was conducted in a series of elegant works (Kim, Wang, Lee, Jang, & Han, 2007; Rubinstein et al., 2008; Yossifon & Chang, 2008).

## 2.3 *Current-voltage response of bipolar systems without EOF*

Finally, we review the $i-V$ of bipolar systems without EOF – the $i-V$ of such systems is given by the dashed purple line in **Figure 2**. There are countless outstanding experimental and numerical works investigating the $i-V$ of these systems – for brevity, we name but a few (Siwy, Heins, Harrell, Kohli, & Martin, 2004; Vlassiouk et al., 2008, 2009; He et al., 2009; Picallo, Gravelle, Joly, Charlaix, & Bocquet, 2013;Ható, Valiskó, Kristóf, Gillespie, & Boda, 2017; Mádai, Valiskó, & Boda, 2019; Lucas & Siwy, 2020; Córdoba, Oca, Dhanasekaran, Darling, & Pablo, 2023) and refer the interested readers to the reviews (Cheng & Guo, 2010; Huang, Kong, Wen, & Jiang, 2018; Pärnamäe et al., 2021) for a complete list. However, there are very few works that have considered the $i-V$ theoretically (reviewed below) – all of which assumed no EOF.

Vlassiouk, Smirnov, and Siwy (2008) first derived a 1D $i-V$ after assuming that the lengths of the positive and negative charged regions were equal, and that the charges of the positive region were minus that of the negative. Their model did not impose that the values of the excess counterion charges, to be defined later in Sec. 3.2, were large or small. Later, Picallo *et al.* (2013) derived an alternative 1D $i-V$ after assuming equal lengths. In contrast to Vlassiouk, Smirnov, and Siwy (2008), they didn't assume the charges were of equal magnitude (and opposite sign), but they did assume that the values of the excess counterion charges were large such that each charged region was ideally selective. **Figure 3** compares the two non-dimensional $i-V$ responses with exact numerical simulations (the normalizations are outlined in Sec. 3.2 below). It can be observed that the $i-V$ of Vlassiouk, Smirnov, and Siwy (2008) holds for all values of $V$ and $N_2$ (defined below), while the $i-V$ of Picallo *et al.* (2013) holds for all voltages, but only for very high values of $N_2$. However, it should be noted that the $i-V$ of Picallo *et al.* (2013) had another advantage – namely, their $i-V$ was also able to account for bulk asymmetric concentrations – that were not accounted for in the other model.

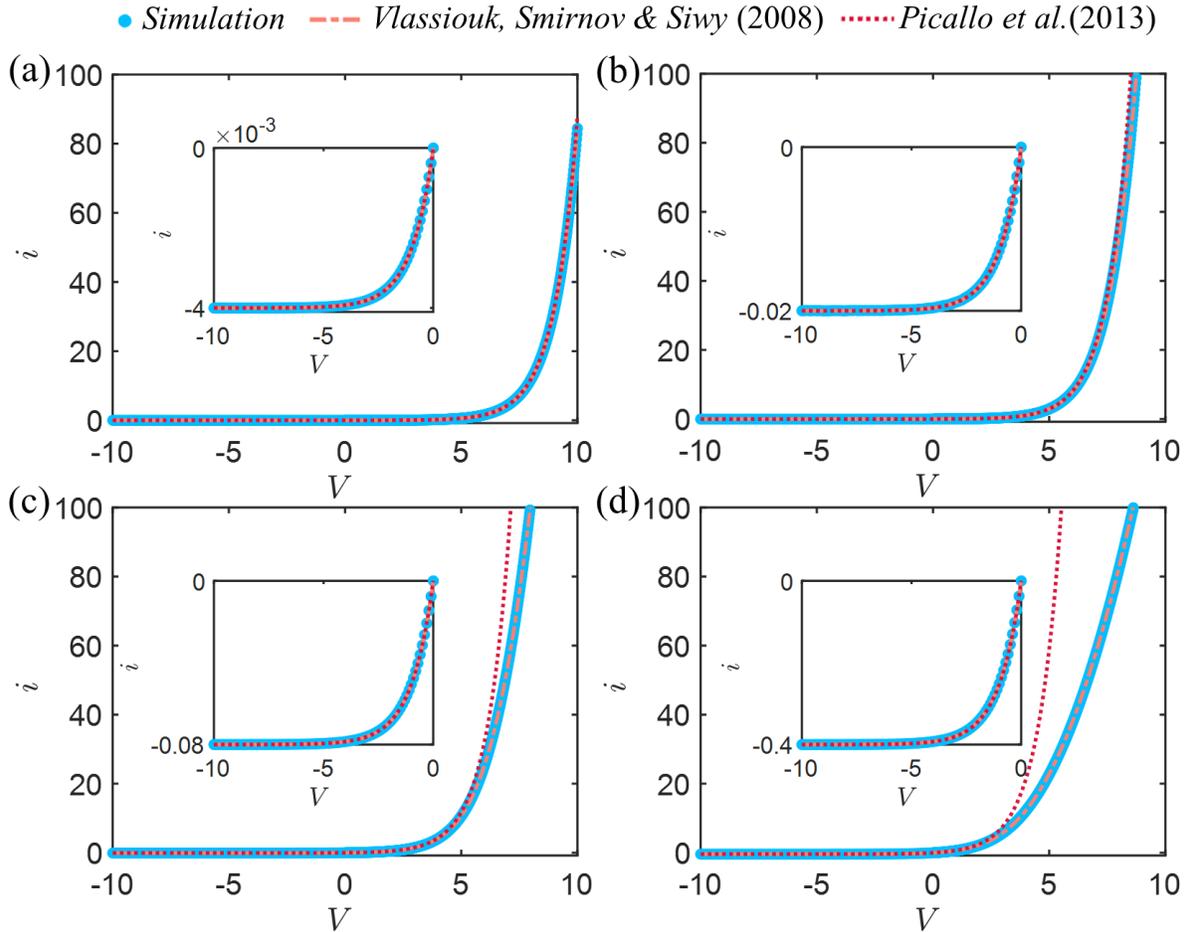

*Figure 3. The (non-dimensional) current(-density)-voltage, $i-V$, curves comparing the two theoretical works of Vlassiouk, Smirnov, and Siwy (2008) and Picallo et al. (2013), and 1D numerical simulations for a bipolar system with varying values of excess counterion charge $N_2$: (a) $N_2 = 2000$, (b) $N_2 = 200$, (c) $N_2 = 50$, and (d) $N_2 = 10$. To allow for a straightforward analysis, here, we have assumed, $L_{2,3} = 1$, $N_2 = -N_3$ and that the effect of the diffusion layers is negligible such that $L_{1,4} = 0$.*

We point out that the $i-V$ of Vlassiouk, Smirnov, and Siwy (2008) was extended later by Green, Edri, and Yossifon (2015), who derived a more general $i-V$ with three changes/modifications. First, their solution considered a 2D geometry. Second, they discovered that one does not need to require symmetry in the geometry and anti-symmetry in the surface charges independently, but rather, there is a more general constraint – this constraint is discussed in Sec. 4 more thoroughly. Third, they accounted for the diffusion layers. This is important because, in any realistic system (**Figure 1**), the bipolar diode is connected to larger reservoirs that always need to be accounted for. Even without EOF they contribute additional resistances. However, when EOF is accounted for, the instability that occurs forms at the interface of the bipolar diode with the diffusion layers.

## 2.4 *Current-voltage response of bipolar systems with EOF*

To date, to the best of our knowledge, there is only one very initial numerical work (Ganchenko, Kalaydin, Ganchenko, & Demekhin, 2018) that has started to address the effects of EOF on the $i-V$ response in bipolar systems. However, in this work, two assumptions are made. One is that the positively and negatively charged regions are anti-symmetrically charged. Second, rather than considering the simplest electrolyte of two species, they consider a four-species electrolyte, which also accounts for water-splitting effects. However, as demonstrated by (Andersen et al., 2012; Nielsen & Bruus, 2014), the physics of a four-species electrolyte in a single diffusion layer is substantially enriched and different from that of a two-species electrolyte. Thus, it is difficult to determine how much the $i-V$ of a four-species electrolyte with EOF varies relative to that of a two-species electrolyte without EOF.

The purpose of this work is to bridge the current knowledge gap. Here, we will precisely consider the $i-V$ response of a bipolar system comprised of two charged regions, where the ratio of the charges is arbitrary, with a two-species electrolyte subject to EOF.

## 3. Problem formulation

This work considers ion transport across a system comprised of four layers: two diffusion layers and two oppositely charged permselective regions. Section 3.1 discusses the 2D geometric setup. Section 3.2 presents the governing equations. Section 3.3 supplements the boundary conditions. Section 3.4 briefly details the numerical methods and simulation parameters. Section 3.5 details various averaging operators repeatedly used in both Part I and Part II.

### 3.1 *Geometry*

**Figure 4** presents a 2D system comprised of four regions of uniform width $W$ and various lengths $L_k$ where the index $k = 1, 2, 3, 4$ denotes each region (top view of **Figure 1**). The two outer regions (Regions 1 and 4) are commonly termed "diffusion layers" – these are uncharged regions filled with an electrolyte. Regions 2 and 3 are negatively and positively charged, respectively, permselective regions. In the following, we will use the $\Delta_k$ notations to denote cumulative lengths within the system

$$\Delta_1 = L_1, \quad \Delta_2 = \Delta_1 + L_2, \quad \Delta_3 = \Delta_2 + L_3, \quad \Delta_4 = \Delta_3 + L_4. \tag{1}$$

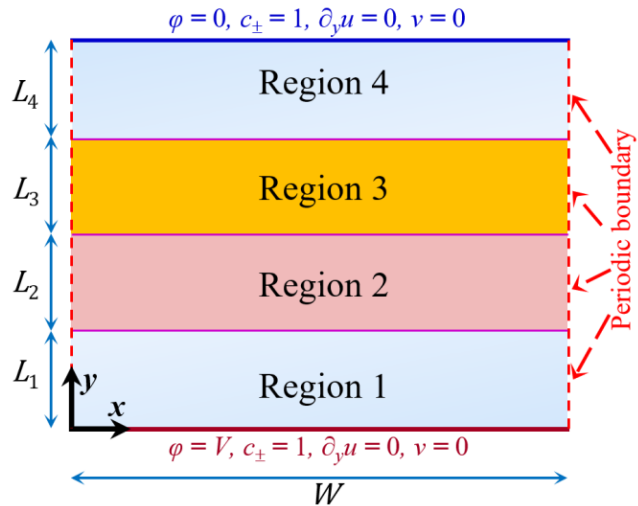

*Figure 4*: *Schematic of a two-dimensional four-layered system comprised of two diffusion layers (Regions 1 and 4) connected by two permselective mediums (Regions 2 and 3). The origin is in the bottom left corner of Region 1. The width of the system is $W$, while the length of each region is given by $L_k$, $k = 1, 2, 3, 4$. At the outer boundary of the diffusion layers, there are two bulk reservoirs with the same electrolyte maintained at an equal, fixed concentration and under a potential drop of $V$ (defined as positive from $y = 0$ to $y = \Delta_4$). The system is periodic at $x = 0$ and $x = W$ (denoted by dashed red lines). All boundary conditions are detailed in Sec. 3.3. We account for EOF in the diffusion layers (Regions 1 and 4), while in the permselective regions (Regions 2 and 3), we do not (see Sec. 3.2).*

### 3.2 Governing equations

The non-dimensional time-dependent equations that govern ion transport through a permselective medium are the Poisson-Nernst-Planck and the Stokes equations. For a symmetric and binary electrolyte ($z_+ = -z_- = 1$) with ions of equal diffusivities ($\tilde{D}_\pm = \tilde{D}$), the non-dimensional equations are

$$\partial_t c_\pm = -\nabla \cdot \mathbf{j}_\pm = \nabla \cdot (\nabla c_\pm \pm c_\pm \nabla \varphi) - \text{Pe}(\mathbf{u} \cdot \nabla) c_\pm, \tag{2}$$

$$2\varepsilon^2 \nabla^2 \varphi = -\rho_e, \tag{3}$$

$$\nabla \cdot \mathbf{u} = 0, \tag{4}$$

$$-\nabla p + \nabla^2 \mathbf{u} + \nabla^2 \varphi \nabla \varphi = 0. \tag{5}$$

Note that, here and in Part II, tilded notations are used for dimensional variables, whereas untilded variables are non-dimensional. Eq. (2) is the Nernst-Planck equation satisfying continuity of ionic fluxes, $\mathbf{j}_\pm$, for cation and anion concentrations, $c_+$ and $c_-$, respectively. The concentrations have been normalized by the bulk concentration $\tilde{c}_0$. The spatial variables have been normalized by a characteristic length $\tilde{L}$ ($\tilde{L}$ can be chosen arbitrarily as one of $\tilde{L}_{1,2,3,4}$ or $\tilde{W}$ so long as one of these lengths is set to unity). Time, $\tilde{t}$, has been normalized by the diffusion time $\tilde{t}_D = \tilde{L}^2 / \tilde{D}$, and the ionic fluxes have been normalized by $\tilde{j}_0 = \tilde{D}\tilde{c}_0 / \tilde{L}$. Equation (3) is the Poisson equation for the electric potential, $\varphi$, which has been normalized by the thermal potential $\tilde{\varphi}_{th} = \tilde{\mathfrak{R}}\tilde{T} / \tilde{F}$, where $\tilde{\mathfrak{R}}$ is the universal gas constant, $\tilde{T}$ is the absolute temperature, and $\tilde{F}$ is the Faraday constant. The parameter $\varepsilon$ is the non-dimensional Debye length,

$$\varepsilon = \frac{\tilde{\lambda}_D}{\tilde{L}}, \quad \tilde{\lambda}_D = \sqrt{\frac{\tilde{\varepsilon}_0 \varepsilon_r \tilde{\mathfrak{R}} \tilde{T}}{2\tilde{F}^2 \tilde{c}_0}}. \tag{6}$$

Herein, $\tilde{\varepsilon}_0$ and $\varepsilon_r$ are the permittivity of vacuum and the relative permittivity, respectively. The non-dimensional space charge density, $\rho_e$ (normalized by $\tilde{F}\tilde{c}_0$), in each of the regions, appearing on the right-hand side of Eq. (3), is given by

$$\rho_e = c_+ - c_- - N_2 \delta_{2,k} - N_3 \delta_{3,k}, \tag{7}$$

where $\delta_{lk}$ is Kronecker's delta and $N_{2,3}$ ($N_2 \geq 0$ and $N_3 \leq 0$) are the non-dimensional volumetric excess counterion charge densities in the permselective regions (normalized by $\tilde{F}\tilde{c}_0$) due to the surface charge densities, $\tilde{\sigma}_s$. In nanochannel systems, this relation is given by $N = -2\tilde{\sigma}_s / \tilde{F}\tilde{c}_0 \tilde{h}$, where $\tilde{h}$ is the height of the permselective region. Equations (4) and (5) are, respectively, the continuity equation for an incompressible solution and the Stokes equation obtained from the full momentum equation after omitting the nonlinear inertia terms due to a small Reynolds number. The non-dimensional velocity vector $\mathbf{u} = u\hat{x} + v\hat{y}$ and pressure $p$ have been normalized, respectively, by a typical velocity $\tilde{u}_0$ and pressure $\tilde{p}_0$,

$$\tilde{u}_0 = \frac{\tilde{\varepsilon}_0 \varepsilon_r \tilde{\varphi}_{th}^2}{\tilde{\mu}\tilde{L}}, \qquad \tilde{p}_0 = \frac{\tilde{\mu}\tilde{u}_0}{\tilde{L}}, \qquad (8)$$

where $\tilde{\mu}$ is the dynamic viscosity of the fluid. Note that the time derivative of the velocity has also been neglected since the Schmidt number [$\text{Sc} = \tilde{\mu}/(\tilde{\rho}\tilde{D})$ where $\tilde{\rho}$ is the mass-density] is large (Rubinstein & Zaltzman, 2015). Correspondingly, the resulting non-dimensional Péclet number, Pe, in Eq. (2), defined as

$$\text{Pe} = \frac{\tilde{u}_0 \tilde{L}}{\tilde{D}} = \frac{\tilde{\varepsilon}_0 \varepsilon_r \tilde{\varphi}_{th}^2}{\tilde{\mu}\tilde{D}}, \qquad (9)$$

is an intrinsic material characteristic of the electrolyte, independent of the bulk concentration or characteristic length. As indicated in Rubinstein and Zaltzman (2000), for a typical aqueous low molecular electrolyte, Pe is of the order of unity (discussed further below).

In this work, we have chosen a non-dimensional analysis that is drastically more robust than the dimensional analysis. For example, reconsider the expression for the excess counterion concentration $N = -2\tilde{\sigma}_s / \tilde{F}\tilde{c}_0\tilde{h}$, which depends on three independent parameters, $\tilde{\sigma}_s, \tilde{c}_0$ and $\tilde{h}$. Any combination of these parameters that yield the same $N$ will have the same result. In these works, we will keep $N_2$ constant and large (such that $N_2 \gg 1$) while we vary $N_3$. This is equivalent to changing the surface charge density in Region 3. Notingly, the scenario $N_3 = 0$ will reduce the system to an ideally selective unipolar system (whose response has been thoroughly investigated).

In this work, we focus on elucidating the effects of the EOI in a bipolar setup. It is essential to realize that the effects of electroconvection manifest themselves differently in the diffusion layers versus the permselective regions. In practicality, the diffusion layers are regions comprised solely of fluids and are dominated by bulk effects. In contrast, the permselective regions are either highly confined nanochannels or nanoporous membranes. In either situation, the effects of the surfaces, through the requirement of no slip and the effects of tortuosity, result in substantially smaller velocities (relative to what appears in the diffusion layers), such that the effect of the velocity is virtually zero. To that end, and to reduce computational costs, we a priorly assume the velocity field in the permselective regions is zero (Rubinstein & Zaltzman, 2015; Abu-Rjal, Rubinstein, & Zaltzman, 2016; Abu-Rjal, Prigozhin, Rubinstein, & Zaltzman, 2017; Ganchenko et al., 2018). In other words, in Regions 1 and 4, we solve Eqs. (2)-(5), while in Regions 2 and 3, we solve only Eqs. (2)-(3).

### 3.3 *Boundary conditions*

The boundary conditions for the closure of the governing equations [Eqs. (2)–(5)] are given below. At the stirred bulk reservoirs, we have a bulk solution with the same electrolyte maintained at an equal, fixed concentration that is experiencing no shear stress and zero inflow into the system and under a potential drop of $V$, (crimson and blue lines located at $y = 0$ and $y = \Delta_4$, respectively, in **Figure 4**)

$$y = 0: \quad c_\pm = 1, \quad \partial_y u = 0, \quad v = 0, \quad \varphi = V, \qquad (10)$$

$$y = \Delta_4: \quad c_\pm = 1, \quad \partial_y u = 0, \quad v = 0, \quad \varphi = 0. \tag{11}$$

At all the internal interfaces, located at $y = \Delta_1, \Delta_2, \Delta_3$ (pink lines in **Figure 4**), we require continuity of the normal ionic fluxes and electric field

$$y = \Delta_1, \Delta_2, \Delta_3: \quad [\mathbf{j}_\pm \cdot \mathbf{n}] = 0, \quad [-\nabla \varphi \cdot \mathbf{n}] = 0, \tag{12}$$

where [...] represents the differences across the interfaces. At the interfaces between the diffusion layers and permselective regions, we impose the common no-slip condition

$$y = \Delta_1, \Delta_3: \quad \mathbf{u} = \mathbf{0}. \tag{13}$$

We complete the boundary conditions by prescribing periodicity at $x = 0$ and $x = W$ (dashed red lines in **Figure 4**)

$$f(0, y, t) = f(W, y, t), \quad \partial_x f(0, y, t) = \partial_x f(W, y, t), \tag{14}$$

where $f$ corresponds to all the variables in Eqs. (2)-(5) [i.e., $c_\pm$, $\varphi$, $\mathbf{u}$ and $p$].

### 3.4 *Numerical simulations.*

Numerical simulations for the 2D four-layer system (**Figure 4**), taking equilibrium as an initial condition, were carried out for the potentiostatic scenario (i.e., constant voltage and time-dependent current) using Comsol. It is important to note that while this work is entirely based on numerical methods, there is nothing novel within the numerical method we have used. Rather, the novelty in this work lies within the results and the analysis, where we demonstrate many new robust phenomena that, to the best of our knowledge, have never been reported. See the Supplementary Material for more details regarding the numerical schemes used to solve these equations, including how the equilibrium initial conditions are computed, parameters, meshing, and more.

### 3.5 *Averaging operator*

To characterize the system response, we utilize both time and spatial averages of any quantity $f$ (e.g., $c_\pm$, $\varphi$, $\rho_e$, and their fluxes, etc.). Here, we define the operators and their notations. The spatial average across the *x*-direction is denoted with an overbar and defined as

$$\overline{f}(y,t) = \frac{1}{W} \int_0^W f(x,y,t) dx. \tag{15}$$

The surface average over any region $k$ is denoted with two overbars and defined as

$$\overline{\overline{f}}_{k=1,4}(t) = \frac{1}{L_{k=1,4}} \int_{L_{k=1,4}} \overline{f}(y,t) dy. \tag{16}$$

To calculate steady state results of either Eqs. (15) or (16), we define the temporal average,

$$\langle \overline{f} \rangle = \frac{1}{T} \int_{t_0}^{t_0+T} \overline{f}\, dt, \quad \langle \overline{\overline{f}} \rangle = \frac{1}{T} \int_{t_0}^{t_0+T} \overline{\overline{f}}\, dt, \tag{17}$$

where $t_0$ is the time at which the system reaches a state where the state is perturbed about a "steady-state" (i.e., a statistically steady-state), and $T$ is the time duration of this "steady-state" (typically the end of the simulations). This average is denoted with angle brackets (chevron brackets).

Of particular importance is the average of the electrical current density, $\tilde{\mathbf{i}} = \tilde{i}_x \hat{\mathbf{x}} + \tilde{i}_y \hat{\mathbf{y}}$. The non-dimensional electrical current density (normalized by $\tilde{F}\tilde{D}\tilde{c}_0 / \tilde{L}$) is defined as

$$\mathbf{i} = \mathbf{j}_+ - \mathbf{j}_- = \text{Pe}(c_+ - c_-)\mathbf{u} - \nabla(c_+ - c_-) - (c_+ + c_-)\nabla\varphi. \tag{18}$$

In the remainder, we will consider the time-dependent $x$-average of the normal component of the electrical current density, $i_y(t)$, at $y = 0$ given by

$$\bar{i}(t) = \frac{1}{W}\int_0^W i_y(x, y=0, t)dx. \tag{19}$$

Another important characterizer of the flow, which will be used throughout this work, is the kinetic energy density (energy per unit volume), $\tilde{E}_k$, in the diffusion layers. The non-dimensional kinetic energy density (normalized by $\tilde{\rho}\tilde{u}_0^2$) is defined as

$$E_k = \tfrac{1}{2}|\mathbf{u}|^2 = \tfrac{1}{2}(u^2 + v^2), \tag{20}$$

and $E_k$ will, often, be subject to any one of the operators given by Eqs. (15)-(17).

Finally, we note that, in general, the $\langle \bar{i} \rangle - V$ response is asymmetric around $V = 0$. The degree of asymmetry is characterized by the rectification factor, RF (Vlassiouk et al., 2008, 2009; Cheng & Guo, 2010)

$$\langle \overline{\text{RF}} \rangle = \left| \frac{\langle \bar{i} \rangle_{V>0}}{\langle \bar{i} \rangle_{V<0}} \right|. \tag{21}$$

In the remainder, we will also make a comparison of the rectification factor. Specifically, we will consider the current mean as well as the current fluctuations. Note that even though $\langle \overline{\text{RF}} \rangle$ is defined by Eq. (21), there is some ambiguity as to how the mean is calculated. In general, one should calculate the mean of the ratio and not the ratio of the means. However, this raises both conceptual difficulties as well as numerical difficulties. To that end, we have found that the ratio of the means is the simplest metric to compare two different states of positive and negative voltages. This definition is loosely related to the concept of error propagation (Sipkens, Corbin, Grauer, & Smallwood, 2023)

$$\sigma_{\overline{\text{RF}}} = \langle \overline{\text{RF}} \rangle \sqrt{\left(\frac{\sigma_{\bar{i}, V>0}}{\langle \bar{i} \rangle_{V>0}}\right)^2 + \left(\frac{\sigma_{\bar{i}, V<0}}{\langle \bar{i} \rangle_{V<0}}\right)^2}. \tag{22}$$

Here we have $\sigma_{\overline{\text{RF}}}$, using the mean current, $\langle \bar{i} \rangle$, and the current density standard deviation, $\sigma_{\bar{i}}$, for positive and negative voltages. Here, it can be noted that the 'total' standard deviation depends on $\langle \overline{\text{RF}} \rangle$ and the mean and standard deviation of each scenario (positive and negative voltages) in an independent manner.

## 4. Steady-state results

In the following, we will present the non-dimensional steady-state response [time- and space average based on Eq. (17)] $\langle \bar{i} \rangle - V$ for three scenarios (unipolar, non-ideal bipolar, and ideal bipolar) without and with EOF.

The unipolar system is a three-layered system that includes only one permselective region. As a result, if both flanking microchannels have the same geometry, the $\langle \bar{i} \rangle - V$ is symmetric $V = 0$. If the flanking microchannels are asymmetric, so too is the $\langle \bar{i} \rangle - V$. These statements hold regardless of whether EOF is included or not. This will be demonstrated shortly.

Bipolar systems, comprised of four layers and including two charged regions of opposite charges, have an inherently more complicated response. In fact, the fourth layer adds an additional degree of freedom, which leads to a parameter that marks the difference between non-ideal and ideal bipolar systems. The parameter that determines the response is a parameter that accounts for the geometric and excess counterion charge density of both permselective regions [see Green, Edri, and Yossifon (2015) for the 2D version of this equation]

$$\eta = \frac{L_3}{L_2} \times \left|\frac{N_3}{N_2}\right| = \begin{cases} 0 & , \text{ unipolar} \\ 0 < \eta < 1, & \text{non-ideal bipolar} \\ 1 & , \text{ ideal bipolar} \end{cases} \qquad (23)$$

It is trivial to see that when either $N_3 = 0$ or $L_3 = 0$, the second charged region does not exist, and the bipolar system reduces to the unipolar system. If, however, there are two charged regions, the ratio $\eta$ will determine the overall response of the system. If $\eta = 1$ (and $L_2 = L_3$), the total excess counterion charges in both regions are equal (but of opposite sign), and the response is what we now term 'ideal' bipolar. If $1 > \eta > 0$, we term the response non-ideally bipolar. The non-ideal bipolar system will exhibit time-dependent and steady-state characteristics of both unipolar and ideal bipolar systems.

Before continuing with our analysis, we wish to make a distinction in the terminology adopted in this work relative to our previous work (Abu-Rjal & Green, 2021). In that work, the $\eta = 1$ scenario was termed a 'symmetric' bipolar system, while the $1 > \eta > 0$ scenario was termed an 'asymmetric' bipolar system. At that time, our choice of using the words 'symmetry' and 'asymmetry' was to emphasize whether there was a symmetry between the total excess counterion charges in both regions. However, $\eta$ does not determine whether the response will be symmetric or not since, <u>in a bipolar system, the response is always asymmetric</u>. Thus, in this work, we see fit to update and change our past terminology such that we now state that $\eta$ determines whether the $\langle \overline{i} \rangle - V$ response is that of an "ideal" bipolar response or a non-ideal bipolar response.

In our numerical simulations, we keep the geometry constant such that $L_{1,2,3,4} = 1$. We set $N_2 = 25$, while varying $N_3$. In doing so, we vary the ratio $|N_3|/|N_2|$. For the sake of simplicity, our initial discussion will focus only on the three scenarios of $N_3 = -25$, $N_3 = -10$, and $N_3 = 0$ corresponding to our three scenarios of ideal bipolar, non-ideal bipolar, and unipolar, respectively. After we have delineated the differences between the three scenarios, we will present a more systematic analysis of the full range of $N_3$ values considered.

Before presenting the comparison of the three scenarios, one last comment is needed. By setting $N_3 = 0$, a four-layered system effectively becomes a three-layered system. To allow for a straightforward comparison of all aforementioned scenarios, we have chosen to keep the overall length of the system the same. Thus, in the unipolar scenario shown in the comparison, we have made the effective length of one of the diffusion layers to $L_3 + L_4 = 2$. This is in contrast to the more standard case of symmetric diffusion layer lengths, where $L_3 + L_4 = 1$. For the sake of completion, we now demonstrate that the qualitative response of these two unipolar systems remains unchanged, with the sole difference being that the system with an asymmetric geometry has a non-unity rectification factor (Green, Edri, & Yossifon, 2015).

**Figure 5**(a) compares the non-dimensional steady-state $\langle \bar{i} \rangle - V$ response between two unipolar systems: symmetric ($L_4 = L_1$, blue lines and circle markers) and asymmetric ($L_4 = 2L_1$, pink lines and square markers). Both responses with EOF exhibit all three current regions described in Sec. 2.2 (and shown schematically in **Figure 2**). The symmetric system has a (anti-)symmetric response around $V = 0$. In contrast, the asymmetric system has an asymmetric response. The degree of asymmetry is quantified through the rectification factor, $\langle \overline{\text{RF}} \rangle$, plotted in **Figure 5**(b). Naturally, for the symmetric case, $\langle \overline{\text{RF}} \rangle \equiv 1$, while for the asymmetric case, our chosen asymmetry leads to $\langle \overline{\text{RF}} \rangle > 1$. While we find that $\langle \overline{\text{RF}} \rangle$ is enhanced by EOF, our numerical simulations show that above $V \simeq 50$, the rectification factor plateaus. This ratio is likely set by geometry – and should be investigated in future works.

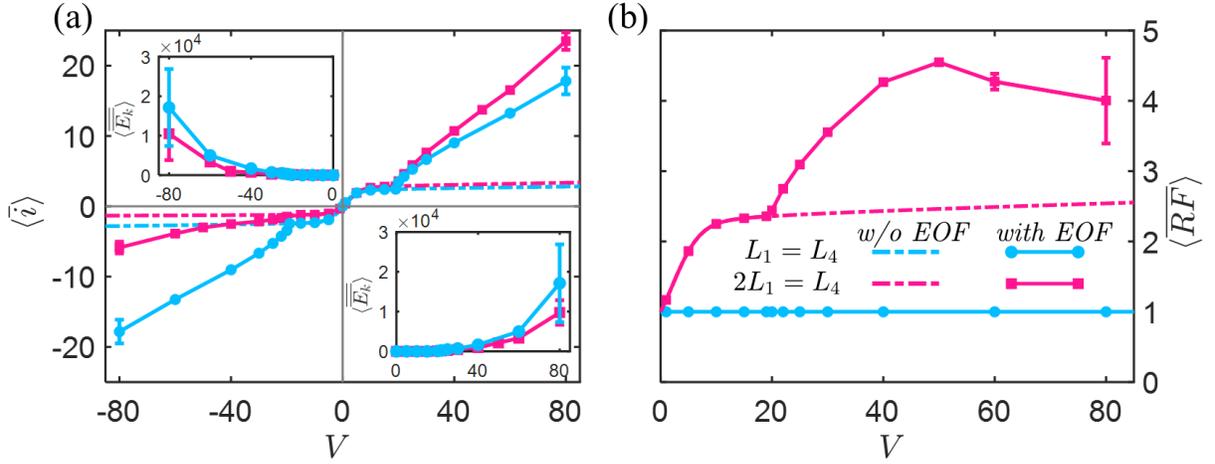

*Figure 5: (a) The (non-dimensional) steady-state current-density-voltage, $\langle \bar{i} \rangle - V$, curves, without and with EOF, for symmetric ($L_1 = L_4$, blue lines) and asymmetric ($2L_1 = L_4$, red lines) unipolar systems [time- and space average based on Eq. (17)]. Note that systems with EOF have the three distinct regions described previously [**Figure 2**]. Insets: The time and surface average of the kinetic energy, $\langle \overline{\overline{E_k}} \rangle$, for negative and positive voltages. (b) The rectification factor versus the voltage. The error bars in (a) denote one standard deviation of the current, while the error bars in (b) denote the standard deviation defined by Eq.* (22).

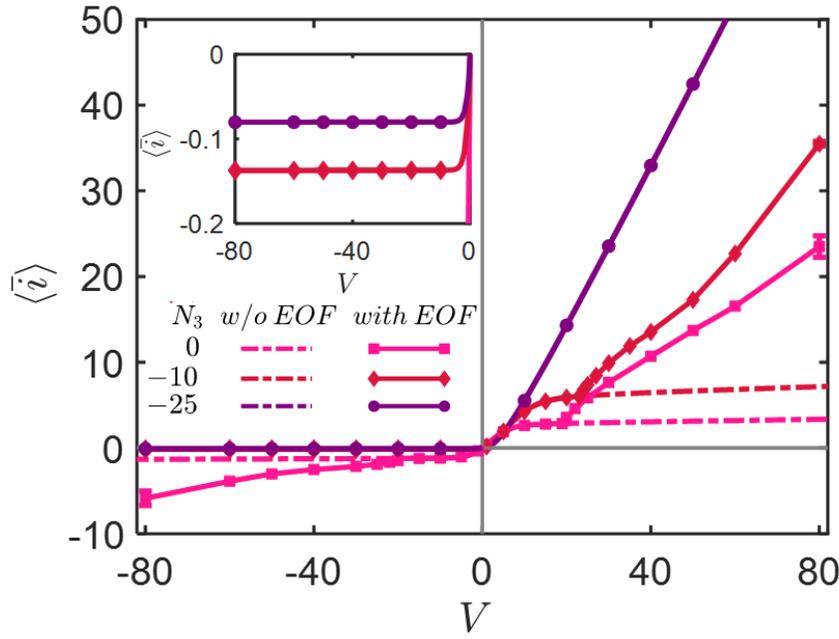

*Figure 6: The (non-dimensional) steady-state current density-voltage, $\langle \bar{i} \rangle - V$, results without EOF (dashed lines) and with EOF (solid lines and markers) for three scenarios: unipolar ($N_3 = 0$), non-ideal bipolar ($N_3 = -10$), and ideal bipolar ($N_3 = -25$) systems. The inset is a zoomed view of the negative voltage near $\langle \bar{i} \rangle = 0$, showing that the $\langle \bar{i} \rangle - V$ curves of bipolar systems do not exhibit OLCs there. The error bars denote one standard deviation of the current.*

**Figure 6** shows the $\langle \bar{i} \rangle - V$ for the three scenarios without EOF (dashed lines) and with EOF (solid lines with markers). For the sake of comparison, here we show the asymmetric unipolar response previously discussed in **Figure 5** and compare it to the two other bipolar scenarios. Notice that the ideal bipolar exhibits a rather remarkable result. The $\langle \bar{i} \rangle - V$ with EOF completely coincides with the $\langle \bar{i} \rangle - V$ without EOF. This remarkable result is due to $\eta = 1$ where the internal "symmetry" of positive and negative charges leads to a "symmetry" in the fluxes. We will further discuss this in the next paragraph. The non-ideal bipolar system exhibits equally remarkable but substantially different behavior. For negative voltages, the $\langle \bar{i} \rangle - V$ with EOF coincides with the $\langle \bar{i} \rangle - V$ without EOF, as in the case of the ideal bipolar system. In contrast, for positive voltages, the $\langle \bar{i} \rangle - V$ with EOF breaks of the $\langle \bar{i} \rangle - V$ without EOF, as in the case of a unipolar system. This, too, will be discussed shortly. Also, we note that for negative voltages, both bipolar systems exhibit limiting currents (even with EOF) that are not observed for the unipolar system with EOF.

To understand why the ideal bipolar scenario with EOF doesn't exhibit a difference relative to the scenario without EOF, we must return to the problem definition in Sec. 3.2 (see last paragraph), where we assumed that the velocity within the permselective material is zero. Essentially, independent of what is occurring within the diffusion layers, we have constrained the response of the bipolar membrane (Regions 2 and 3) to be identical to the response of a bipolar membrane in the convectionless scenario. This is immensely important since the $\langle \bar{i} \rangle - V$ for the $\eta = 1$ scenario (discussed in Sec. 2.3 and shown in **Figure 2**) assumes that the steady-state salt current density, $j = j_+ + j_-$, is zero ($j = 0$) such that $j_+ = -j_-$ (Green, Edri, & Yossifon, 2015). From the steady-state point of view, we realize that by virtue of the internal symmetry, for every positive ion transported from Region 1 to Region 4, there is a negative ion transported from Region 4 to Region 1. These counter fluxes are responsible for stabilizing the extended space charge layer that forms in each of the regions. The convection-less dynamics have been discussed thoroughly in our past work (Abu-Rjal & Green, 2021), and the dynamics with convection will be discussed in Part II. However, if one removes the $\eta = 1$ condition, then $j \neq 0$. Thus, non-ideal bipolar systems do exhibit over-limiting currents.

Our last comment pertaining to **Figure 6** is with regard to negative voltages. For a unipolar system, over-limiting currents are observed, while for bipolar systems, they are not. This is because for unipolar systems, there is no inherent difference between the two diffusion layers and the switch of $V > 0$ with $V < 0$ (or vice-versa, see **Figure S4** in Supplementary Material). This is in contrast to bipolar systems, where the internal (and inherent) electric field from the positive region (here, Region 2) to the negative region (here, Region 3) leads to a change in the behavior of the dynamics of the diffusion layers when $V > 0$ and $V < 0$ are switched. Here, too, the convection-less dynamics have been discussed thoroughly in our past work (Abu-Rjal & Green, 2021), and the dynamics with convection will be discussed in Part II.

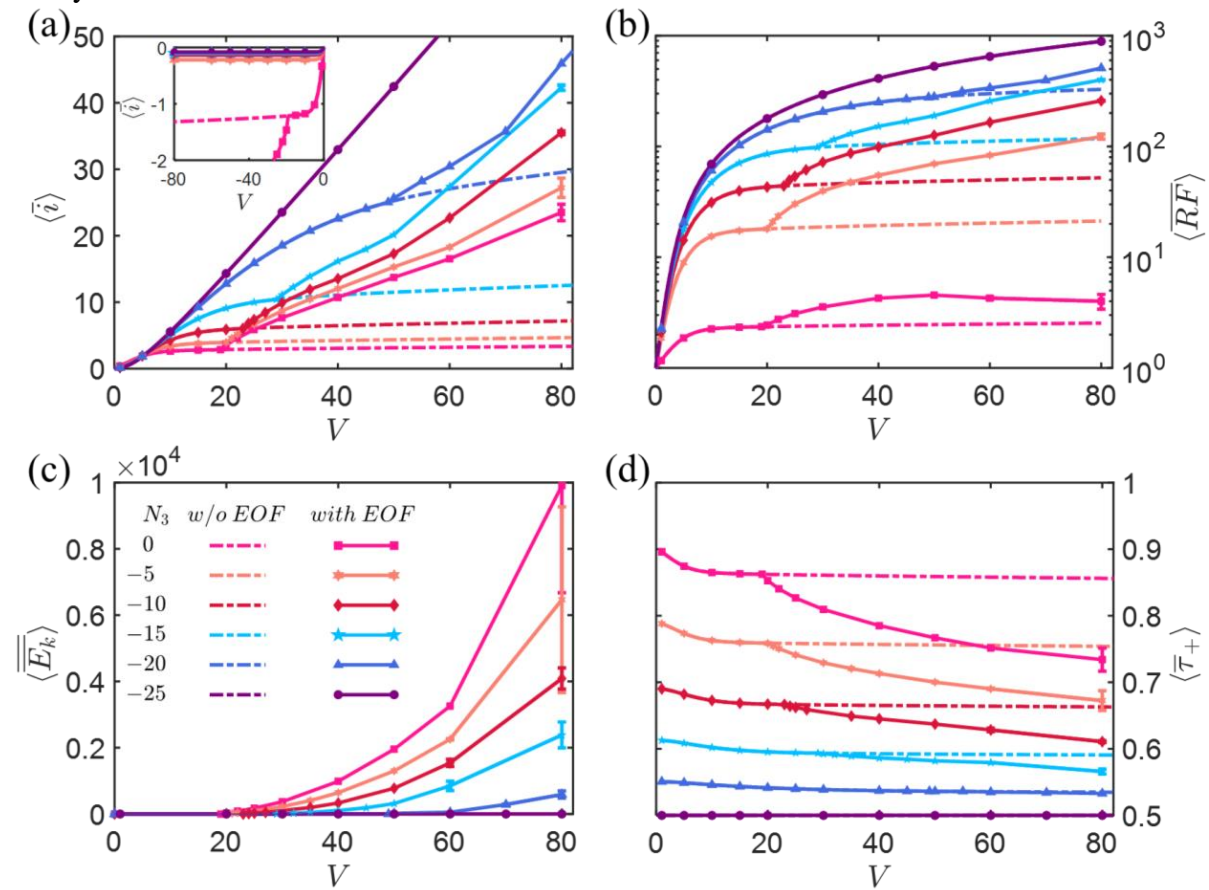

*Figure 7*: *The (non-dimensional) steady-state voltage-dependent results for (a) the current density-voltage response, $\langle \bar{i} \rangle - V$, (b) a semilog$_{10}$ plot of the rectification factor, (c) the surface*

*average of the kinetic energy, $\left\langle \overline{\overline{E_k}} \right\rangle$, in Region 1, and (d) the cationic transport number, $\left\langle \overline{\tau}_+ \right\rangle$, for the scenarios without EOF (dashed lines) and with EOF (solid lines and markers) for several values of $N_3$. The inset of (a) is the $\left\langle \overline{i} \right\rangle - V$ for negative voltages. The vertical bars denote one standard deviation error bar.*

Finally, **Figure 7** presents a thorough scan for varying values of $N_3$ (and, thus $\eta$). **Figure 7**(a) shows the steady state $\left\langle \overline{i} \right\rangle - V$ for several values of $N_3$ without EOF (dashed lines) and with EOF (solid lines). We vary $N_3$ from a unipolar scenario ($N_3 = 0$ given by the pink line) and an ideal bipolar scenario ($N_3 = -N_2 = -25$ given by the purple line). Save for some quantitative differences, all the non-ideal diodes behave similarly to the one discussed in **Figure 6**. The changes in all the steady-state $\left\langle \overline{i} \right\rangle - V$ responses can be correlated to the steady-state kinetic energy [**Figure 7**(c)]– the larger the kinetic energy, the stronger the mixing is, and with it comes an increased current [**Figure S5** in Supplementary Material shows that as the voltage drop is increased, the vortex array loses its stable structure, and a more chaotic and efficient mixing takes place – a similar result was shown by (Druzgalski, Andersen, & Mani, 2013)]. Supplementary Material **Figure S6** shows that the critical voltage, $V_{cr}$, varies as $N_3$ is varied. Two interesting limits are noteworthy. For the ideal bipolar response, the voltage is substantially increased. For the unipolar scenario, the critical voltage is the smallest (~ 20). This value is similar to the linear stability analysis prediction by (Zaltzman & Rubinstein, 2007; Demekhin, Nikitin, & Shelistov, 2013; Abu-Rjal, Rubinstein, & Zaltzman, 2016) and those found in the numerical simulations of (Demekhin, Nikitin, & Shelistov, 2013; Druzgalski, Andersen, & Mani, 2013). Future works should follow the works of (Zaltzman & Rubinstein, 2007; Demekhin, Nikitin, & Shelistov, 2013, 2014), and undertake a linear stability analysis to investigate the critical voltage. One last comment regarding **Figure 7**(c) is needed. **Figure S7** shows that we can fit the kinetic energy to be a parabolic profile such that $\left\langle \overline{\overline{E_k}} \right\rangle = \alpha(N_3)[V - V_{cr}]^2 + \beta(N_3)$. Here, $\alpha(N_3)$ and $\beta(N_3)$ are fitting parameters that depend on $N_3$ (and possibly $N_2$). We are currently unable to rationalize this result – and perhaps this, too, will be resolved within the future work that conducts the linear stability analysis.

Unsurprisingly, as the kinetic energy increases, so does the rectification factor [**Figure 7**(b)]. However, the increases in the currents and the rectification factor are not without consequences. From **Figure 7**(d), we can observe that the transport number

$$\tau_+ = \frac{j_+}{j_+ - j_-} = \frac{1}{2}\left(1 + \frac{j}{i}\right), \tag{24}$$

which is a proxy for the overall selectivity of the channel (Abu-Rjal, Chinaryan, Bazant, Rubinstein, & Zaltzman, 2014; Green, Abu-Rjal, & Eshel, 2020), decreases with decreasing $N_3$ and with the inclusion of EOF. In a unipolar system, when $\tau_+ \equiv 1$, the system is ideally selective, while $\tau_+ \equiv 1/2$ corresponds to a vanishingly selective system. However, such a definition is not simple for bipolar systems, where one of the components is ideally selective, and the other is not [see Sec. 4.2 of Abu-Rjal & Green (2021) for a thorough discussion on the permselective capability of each region separately versus that of the entire system]. The reduction in the transport number, which could be detrimental to desalination, should be considered in future works.

Before proceeding to the discussion, we return to our previous statement (given in Sec. 1) that most nanoporous materials and nanochannels behave the same. This statement is true in so far as all the major transport characteristics remain unchanged. Namely, the Ohmic conductance and the limiting currents predicted for all systems are rather independent of whether the system is 1D, 2D, or 3D. In contrast, we expected that the over-limiting current response of the system will vary as the geometry and the many boundary conditions are varied. This statement should be somewhat intuitive. First, it has been shown that the electroconvective instability is incapable of forming in a purely 1D system (Rubinstein & Zaltzman, 2000). However, if one then considers a 2D system, where the width of the diffusion layers and the permselective materials are the same, the instability does appear above critical voltages. Such a 2D system is quasi-1D at low voltages but 2D at over-critical voltages. A similar statement holds for 3D systems. All systems in which the diffusion layers and the permselective materials have the same height and width are termed "homogeneous" systems. Except for the fact that 2D and 3D homogeneous systems can predict the instability, they can be made to further differ from pure 1D systems in the following manner. In our work, as in most works, one typically assumes that the system is periodic [our Eq. (14)]. Such an assumption is tantamount to assuming that the width and length of the system are infinite and that the effects of the outer boundaries are negligible. However, this does not need to be the case. For example, if at $x = 0, W$, there are walls, the periodic BC should be replaced with no-flux ($\mathbf{j}_\pm \cdot \mathbf{n} = 0, -\nabla \varphi \cdot \mathbf{n} = 0$) and no-slip ($\mathbf{u} = 0$) BCs. The past work of Abu-Rjal et al. (2019) on unipolar systems has shown that the "overall" response of the system does not change in the following manner: over-limiting currents still appear. However, both the time-transient and steady values of this scenario varied slightly relative to the purely homogeneous state. In changing the BC from an infinite system to a finite-size system, the effects of "confinement" have been introduced. In general, we believe that so long as the height and width of homogeneous systems are large enough, then the robustness of the EOI will be the primary determinant of the system, and the response will not vary substantially.

There is yet another way to introduce the effects of confinement into the system. Consider systems in which the permselective material interface is smaller than the interface of the diffusion layers (i.e., a simple nanochannel system). These are called "heterogeneous" systems. The introduction of heterogeneity also requires the change of boundary conditions at the permselective interface [some regions will utilize those given in Eq. (12), while others will require no flux]. While the effects of heterogeneity in overlimiting conditions are not fully understood, they are understood in underlimiting conditions. Several works have shown that the Ohmic conductance and limiting currents can be theoretically predicted (Rubinstein, 1991; Rubinstein, Zaltzman, & Pundik, 2002; Green & Yossifon, 2014; Green, Shloush, & Yossifon, 2014). These underlimiting predictions were later confirmed experimentally (Green, Park, & Yossifon, 2015; Green, Eshel, Park, & Yossifon, 2016). While these models have been very useful for predicting the underlimiting response and some of the trends observed for overlimiting responses, they are not capable of explaining the measurements of a confined heterogeneous system in which the critical voltage for over-limiting current increases as the degree of confinement is increased (Yossifon, Mushenheim, & Chang, 2010). Several numerical models focused on this problem (Schiffbauer, Demekhin, & Ganchenko, 2012; Andersen, Wang, Schiffbauer, & Mani, 2017), in which a Hele-Shaw term was added to account for this confinement. However, these models typically assume that the system is a confined homogeneous system.

This work has considered the scenario of a 2D unconfined periodic system in a bipolar system. Future works should focus on increasing the level of complexity by considering either non-periodic BCs, or heterogeneous systems, or both. We reiterate that depending on the scenario, we expect that the overall robustness of the response of the system will remain unchanged. However, we also expect that there will be conditions in which the response changes drastically.

## 5. Discussion and summary

This work focuses on elucidating the behavior of the $\langle \bar{i} \rangle - V$ response of ideal and non-ideal bipolar systems subjected to EOI. To highlight the surprising and remarkable results uncovered in our simulations of the bipolar system, we first leveraged our understanding of the $\langle \bar{i} \rangle - V$ response of the more investigated unipolar system. There, it has been known for quite some time that the EOI is responsible for OLCs. We also utilized our understanding from our previous work (Abu-Rjal & Green, 2021) on how the $\langle \bar{i} \rangle - V$ response of ideal and non-ideal bipolar without EOF varies as $\eta$ is varied. We then simulated several ideal and non-ideal bipolar systems for a wide range of $V$ and $\eta$. Our findings have led to several surprising results, which can be divided into two: ideal and non-ideal bipolar systems:

- In ideal bipolar systems, with $\eta = 1$, the scenarios without EOF and with EOF have the same steady-state $\langle \bar{i} \rangle - V$ responses.
- In non-ideal bipolar systems, with $1 > \eta > 0$, the $\langle \bar{i} \rangle - V$ shares some of the characteristics of both unipolar and pure bipolar systems. For example, when $V > 0$, over-limiting currents are observed, while for $V < 0$ limiting currents are observed.

While this work further advances our knowledge of ion transport with EOI in bipolar systems, it also underscores that an in-depth understanding of the dynamics involved remains lacking, and further investigations are needed. Fortunately, the community has most of the required theoretical and numerical tools to undertake this task of delineating the various parameters and mechanisms that control the response of the $\langle \bar{i} \rangle - V$. This work, hopefully, will provide the roadmap needed for such a vast undertaking.

**Acknowledgments.** We thank Lyderic and Marie-Laure Bocquet, and their groups for enlightening discussions regarding the ideal bipolar steady-state current-voltage response during YG's visit to their labs. We also thank Oren Lavi for several in-depth discussions and very useful comments on how to streamline the results of this two-part work.

**Competing interests.** The authors declare no conflict of interest.

**Funding statement.** This work was supported by Israel Science Foundation grants 337/20 and 1953/20. We acknowledge the support of the Ilse Katz Institute for Nanoscale Science & Technology.

**Data Availability Statement.** Raw data will be made available upon request from the corresponding author (Y.G.).

**Supplementary Material and Movies.** Supplementary material documents and supplementary movies intended for publication have been provided with the submission. These are available at https://doi.....

## Appendix A. Glossary of terminology

Here, we provide a brief glossary of terminologies used throughout this Part and Part II.

*Cations*: positive ions, $c_+$.

*Anions*: negative ions, $c_-$

*Counterions*: Ions with an opposite charge to that of the surface charge in the selective region.
*Coions*: Ions with a similar charge to that of the surface charge in the selective region.
*Permselectivity*: A symmetry-breaking property, due to the surface charge density, that allows for the preferential passage of counterions over coions.
*Ion-selective medium*: A material that exhibits permselectivity (in this work, Regions 2 and 3).
*Excess counterion concentration*: The additional concentration of counterions present within a permselective material (due to the fixed surface charge). Here, $N_{2,3}$.

*Diffusion layer*: Regions that are uncharged and where ion transport is dominated by diffusion (in this work, Regions 1 and 4).

*Unipolar system*: A three-layer system consisting of only one ion-selective region [either negatively ($N_3 = 0$) or positively ($N_2 = 0$) charged] flanked by two diffusion layers.

*Symmetric unipolar system*: A unipolar system with geometrically symmetric diffusion layers, $L_1 = L_4$. In this work, our unipolar system occurs when $L_3 = 0$ and $N_3 = 0$.

*Asymmetric unipolar system*: A unipolar system with geometrically asymmetric diffusion layers, $L_1 \neq L_4$. In this work, $L_1 = L_3 + L_4$.

*Bipolar system*: A four-layer system consisting of two oppositely charged ion-selective regions ($N_{2,3} \neq 0$ and $L_{2,3} \neq 0$) flanked by two diffusion layers.

*Ideal-bipolar system*: A bipolar system wherein the two ion-selective regions have the same absolute value of the product of geometry and excess counterion concentration, $|L_2 N_2| = |L_3 N_3|$. In our previous work (Abu-Rjal & Green, 2021), we termed this a "symmetric bipolar system".

*Non-ideal bipolar system*: A bipolar system where the two permselective regions have different absolute values of the product of geometry and excess counterion concentration, $|L_2 N_2| \neq |L_3 N_3|$. In our previous work (Abu-Rjal & Green, 2021), we termed this an "asymmetric bipolar system".

**Author ORCIDs.**
Ramadan Abu-Rjal https://orcid.org/0000-0002-1534-9710
Yoav Green https://orcid.org/0000-0002-0809-6575